# Isotope Effect on the Superconducting Transition Temperature of Na$_x$CoO$_2 \cdot y$H$_2$O

Mai Yokoi, Yoshiaki Kobayashi, Masatoshi Sato and Shunji Sugai

*Department of Physics, Division of Material Science, Nagoya University, Furo-cho, Chikusa-ku, Nagoya 464-8602*

The effect of the $^{16}$O→$^{18}$O substitution on the superconducting transition temperatures $T_c$ of Na$_x$CoO$_2 \cdot y$H$_2$O has been studied for several sets of samples. Each set has two kinds of samples, one with the oxygen sites in the CoO$_2$ layers filled with $^{16}$O atoms and another with these sites partially substituted with $^{18}$O isotope. These two kinds of samples were prepared in parallel in whole preparation processes. For all the sample sets, we have not found any significant difference of the $T_c$ values between the non-substituted and substituted samples, which indicates that the oxygen isotope effect is not appreciable in Na$_x$CoO$_2 \cdot y$H$_2$O. The result, however, does not necessarily exclude the possibility of the phonon mechanism of the superconductivity, because the strong correlation effect suppresses the isotope effect as pointed out previously.

## 1. Introduction

Na$_x$CoO$_2 \cdot y$H$_2$O ($x \sim 0.35$, $y \sim 1.3$) has the triangular lattice formed of edge-sharing CoO$_6$ octahedra and exhibits superconductivity with the transition temperature $T_c \sim 4.5$ K.[1] Due to the structural characteristic suggesting the frustrated nature of the electrons, the superconductivity of this system was first expected by many researchers to be unconventional. Then, the spin-triplet pairing was proposed rather strongly by considering the possible characteristic of the Fermi surface topology with six hole pockets of the $e_g$' band near the K points in the reciprocal space.[2-4]

We have carried out various kinds of experimental studies and revealed by measuring, in particular, the $^{59}$Co-NMR Knight shifts that the spin-singlet pairing is realized in this system.[5-7] We have also found that the rate of the $T_c$-suppression by non-magnetic impurities is small.[8] These results are simply explained by considering the $s$-wave superconducting pairs.

On the other hand, the coherence peak has not been observed appreciably in the $T$-dependence of the $^{59}$Co-NQR relaxation rate $1/T_1$,[9-11] suggesting that the anisotropic (probably $d$-wave) order parameter is realized.

From the theoretical side, the possibility of the $s$-wave pairing by the ordinary electron-phonon interaction has been discussed by Yada and Kontani,[12] where the shear and breathing phonons (in-plane $E_{1g}$ and out-of-plane $A_{1g}$ vibrations of oxygen atoms in the CoO$_2$ layer, respectively) which strongly couple with the $t_{2g}$ electrons are considered. The extended $s$-wave superconductivity by spin- and charge-fluctuations is also suggested by Kuroki *et al.*[13] The $d$-wave superconductivity is also expected by simply considering the electron pairing mediated by the spin fluctuations, which are expected for the electrons with frustrated nature.[14-17]

Here, in order to get see what excitation is important for the electron pairing, the oxygen isotope effect on $T_c$ has been studied, and no appreciable effect of the substitution of $^{16}$O with $^{18}$O has been found. In this paper, we argue what the result indicates.

## 2. Experiments

Samples of Na$_x$CoO$_2$ ($x \sim 0.75$) used in the isotope substitution were prepared by solid reactions: Mixture of Na$_2$CO$_3$ and Co$_3$O$_4$ were pre-heated at 750 °C for 12 h, and these samples were pressed into pellets and annealed at 830 – 850 °C for 24 h. These pellets were pulverized or crushed into grains with the typical sizes of 4×4×1 mm$^3$. Then, the pulverized sample was divided into a set of two, each of which weighs 0.2 g (set A). The crushed grains were divided into three sets of two samples (sets B-D), each sample of which also weighs 0.2 g. The substitution of $^{16}$O with $^{18}$O isotope was performed by using two samples belonging to the same set, where the whole processes were carried out in parallel to guarantee that the preparation conditions were equal: One and another samples belonging to a same set were wrapped separately in platinum sheets and sealed in quartz tubes filled with $^{16}$O$_2$ and $^{18}$O$_2$ (97 % enriched), respectively, where the molar number of the filled O$_2$ gas was estimated to be ~2×10$^{-3}$) mol. Then, the tubes were annealed simultaneously in a same furnace at 830 °C for 12h. The heat treatment with fresh $^{18}$O$_2$ gas was carried out 3 times to achieve the high exchange rate. (If the gas exchange is homogeneously and completely carried out, the final fraction of the $^{18}$O occupation should be 0.90.) The obtained samples were found to be the single phase.

Although we tried to estimate the isotope exchange rate by measuring the weight changes during the heat treatment, the significant weight loss possibly due to the Na evaporation made it difficult to precisely estimate the exchange rate. However, we found that for each set, the weight of the sample with $^{18}$O isotope was larger than that of $^{16}$O sample by a similar amount to those found for the other sets. It guarantees that the isotope exchange was carried out for all sets.

To estimate the isotope exchange rate, Raman scattering experiments were performed for set B. In the studies, the exchange rate at flat outer-surface of the crushed grain was estimated to be close to ~100 %. (However, note that because the $^{18}$O gas 97 % enriched, the exchange rate is less than the value.) We measured the exchange rate inside of the grains, too, by using the cleaved surfaces and estimated as ~50 %.

Then, the pulverized samples annealed in $^{16}$O or $^{18}$O were immersed into the $Br_2/CH_3CN$ solution for several days to de-intercalate Na ions, and washed with distilled water. After these processes, superconducting powder samples of $Na_xCoO_2 \cdot yH_2O$ were obtained. These samples were kept at room temperature in the atmosphere of 100 % humidity.

The superconducting diamagnetism was measured with a SQUID magnetometer (Quantum Design) in the conditions of the zero-field cooling (ZFC) and field cooling (FC) with the external field $H = 5$ G. The magnetic susceptibility $\chi$ was also measured with $H = 1$ T between 2 and 260 K.

### 3. Results and Discussion

The $T$-dependences of the magnetic susceptibilities $\chi$ of the samples of $Na_xCoO_2 \cdot yH_2O$ with and without $^{18}$O substitutions (set A) are shown in Fig. 1. They were measured with the magnetic field of 1 T. The $\chi$-$T$ curves of the two kinds of samples overlap almost completely, suggesting that qualities of these samples are almost equal. For the other sets of samples prepared in parallel, we have found similar results as that of set A.

For the estimation of the isotope exchange rate, Raman scattering experiments were carried out at room temperature for the samples of set B of $Na_xCoO_2$ ($x \sim 0.75$) before the $H_2O$ intercalation. Cleaved surfaces were used for both of the $^{18}$O and $^{16}$O samples to study the inside parts of the samples. Outer-surface part of the grains was also used for $^{18}$O sample. The data are shown in Fig. 2. For the $^{16}$O samples, two strong phonon modes have been observed at around 465 cm$^{-1}$ and 577 cm$^{-1}$, which can be identified to be the in-plane $E_{1g}$ and out-of-plane $A_{1g}$ modes of oxygen atoms in the $CoO_6$ octahedra, respectively.[18] If the substitution with $^{18}$O is complete, the peak positions of these $E_{1g}$ and $A_{1g}$ modes are expected to shift to the lower energy side by a factor of ~ $(M_{16}/M_{18})^{1/2}$, with $M_{16}$ and $M_{18}$ being the atomic weights of $^{16}$O and $^{18}$O, respectively. Actually, the signals from the outside surface of the grains of the $^{18}$O sample were observed at the positions close to the values (~ 438 cm$^{-1}$ and ~ 544 cm$^{-1}$ for the in-plane $E_{1g}$ and out-of-plane $A_{1g}$ modes, respectively) expected for the almost 100 % exchange. On the other hand, from the cleaved surface of the $^{18}$O samples corresponding to its inner parts, signals at ~ 451 cm$^{-1}$ for the in-plane $E_{1g}$ mode and at ~ 558 cm$^{-1}$ for the $A_{1g}$ mode were observed. Here we use the positions of the center of gravity, because the peak is slightly broadened. The isotope exchange rate at the inner part is estimated to be about 50 %, indicating that the equilibrium of $^{16}$O and $^{18}$O has not been reached over the sample volume.

In Fig. 3(a), the magnetic susceptibilities $\chi$ of the samples of set A of $Na_xCoO_2 \cdot yH_2O$ measured with $H = 5$ G under the conditions of ZFC and FC are shown. Figure 3(b) shows the data of $\chi$ measured around $T_c$ at a rather fine step of $T$ under the ZFC. The $T_c$ values were defined in two ways. One is defined as the temperature of the intersection of two straight lines fitted to the $\chi$-$T$ data in the $T$ regions around 4.8 K ($> T_c$) and around the $T$ point where $4\pi\chi$ has the value of -0.005 (see Fig. 3(b)). Another is defined as the onset temperature of $\chi$-decrease with decreasing $T$. Although the latter definition of $T_c$ is slightly ambiguous, we use them as the rough estimations. These studies were also carried out for sets B, C and D, and obtained $\chi$-$T$ curves are shown in Fig. 4. The $T_c$ values of $^{16}$O and $^{18}$O samples are shown for all the sets in Table I, where, $T_c$ values estimated as the cross point of the two fitted lines are shown by $T_c(^{16}O)$ and $T_c(^{18}O)$ for the $^{16}$O and $^{18}$O samples, respectively, and the onset temperatures are shown by $T_c(^{16}O$ onset) and $T_c(^{18}O$ onset).

If $Na_xCoO_2 \cdot yH_2O$ samples are kept at room temperature, their $T_c$ values change, as reported in ref. 19, with time $t$ elapsed after the preparation. This $t$ dependence of $T_c$ can be well described as the $\nu_{Q3}$ dependence,[20-22] where $\nu_{Q3}$ ($\cong 3\nu_Q$) is the $^{59}$Co NQR frequency of the $\pm 5/2 \leftrightarrow \pm 7/2$ transition, Here, we have studied the $T_c$ values of sets A and C by changing the time $t$ kept at room temperature to ensure that this $t$ dependence does not introduce any confusion into the oxygen isotope effect on $T_c$. The $t$ dependences of the $T_c$ values are shown in Fig. 5. For set A, the $T_c$ values decrease gradually with increasing $t$ for both the $^{18}$O and $^{16}$O samples. At a given $t$, the $T_c$ values of the $^{18}$O and $^{16}$O samples are almost equal in the $t$ region between 0 and 620 h. For $t > 620$ h, $T_c$ of the $^{16}$O sample becomes slightly lower than that of the $^{18}$O sample. At $t = 970$ h, $T_c(^{18}O)$ is 0.07 K higher than $T_c(^{16}O)$, while $T_c(^{18}O$ onset) is 0.04 K higher than $T_c(^{16}O$ onset). For set C, the $T_c$ values of the two kinds of samples at a given $t$ are almost equal in the $t$ region between 0 and 890 h. In the lower panel of Fig. 5, the $t$-dependences of $-4\pi\chi$ at $T = 1.9$ K are shown for sets A and C. They decrease with $t$. The slightly lower $T_c$ values of the samples of set C than those of set A, found at any fixed time $t$, may be related to this small values of $-4\pi\chi$. In any case, nearly the same $t$ dependences of the $T_c$ values found in each set of the substituted and non-substituted samples guarantee that the oxygen isotope effect can be properly estimated by comparing the $T_c$ values of the two kinds of samples in each set.

We have so far presented experimental data for the four sets of $^{16}$O and $^{18}$O samples. Now, if we consider that the electron-phonon interaction is relevant to the superconducting pair formation, and if we simply expect that $T_c$ is lowered by a factor of $(M_{16}/M_a)^{1/2}$, with $M_a$ being the averaged oxygen atomic mass, the difference of $T_c$ between $^{18}$O and $^{16}$O samples $\Delta T_c$ ($\equiv T_c(^{18}\text{O}) - T_c(^{16}\text{O})$) is expected to be ~ -0.13 K for the 50 % substitution (inner parts) and ~ -0.25 K for the almost complete substitution (outer surface parts). The value of $\Delta T_c$ (~ 0 ± 0.03 K) observed in the present study is very small as compared with the expected value. We stress that the smallness of $\Delta T_c$ is reproduced for all the sets prepared here, even though the preparation processes is not very simple. It is also stressed that the careful studies on the $t$ dependence of $T_c$ have removed the ambiguity of the values of $\Delta T_c$, which may originate from the different $t$ dependence of $T_c$ between non-substituted and substituted samples. Then, we can conclude that the isotope effect on $T_c$ is almost inappreciably small in the present system. The effect of the inhomogeneity of the $^{16}$O→$^{18}$O exchange rate does not change this conclusion: If $\Delta T_c$ is meaningfully large, the surface parts of the grains of the substituted samples have the lower $T_c$ than the inner parts. Therefore, even in the measurements with the ZFC condition, the superconducting diamagnetism should be smaller for the substituted samples than for the $^{16}$O sample in the $T$ region, at least, just below the superconducting transitions.

Now, what the above result indicates? Does it restrict the superconducting mechanism only within the interactions among the electrons? However, the situation is not so simple, because there are several example systems such as Ru and Zr, in which the isotope effect is small due to the strong Coulomb interaction,[23] even though they exhibit the phonon mediated superconductivity. On this point, Yada and Kontani have discussed the isotope effect on $T_c$ for the present system, showing that the values of $\Delta T_c$ becomes very small with increasing Coulomb interaction $U$, even if the superconductivity is caused by the electron-phonon interaction.[24] Using their results, $\Delta T_c$ is expected to be ~ 0.02 K for the complete exchange rate, $U$ = 5 eV and $\Delta$ = -0.025 eV, where $\Delta$ is the distance between the Fermi level and the top energy of the $e_g$' band located below the Fermi level. (This band structure is consistent with the experimental result that the hole pockets of the $e_g$' band near the K points in the reciprocal space.[25])

Anyhow, we have experimentally shown that the oxygen isotope effect is very small in the present system.

## 4. Summary

Studies of oxygen isotope effect on $T_c$ of Na$_x$CoO$_2$·$y$H$_2$O have been carried out, where the smallness of the $T_c$ change upon the O$^{16}$ → O$^{18}$ substitution is shown in the reproducible way for the four sets of the $^{18}$O-substituted and non-substituted samples. In the studies, by measuring the $T_c$ values as functions of the time elapsed after the preparation, we have confirmed that the $t$ dependences of the $T_c$ values of the $^{18}$O-substituted and non-substituted samples are very similar and do not introduce the ambiguity of $\Delta T_c$ values.

The smallness of $\Delta T_c$ does not exclude the electron-phonon mechanism of the superconductivity of this system, because the strong Coulomb interaction can explain the small values. Of course, the result does not exclude the electron-electron mechanism such as the spin fluctuation mediated superconductivity within the condition of the singlet pairing.

Acknowledgments –The work is supported by Grants-in-Aid for Scientific Research from the Japan Society for the Promotion of Science (JSPS) and by Grants-in-Aid on priority area from the Ministry of Education, Culture, Sports, Science and Technology.

**Figure caption**

Fig. 1. The magnetic susceptibilities $\chi$ measured with the magnetic field $H$ of 1 T are shown against $T$ for the $^{18}$O and $^{16}$O samples of Na$_x$CoO$_2 \cdot y$H$_2$O (set A). They were taken under the zero-field cooling (ZFC) condition.

Fig. 2. Data of the Raman scattering taken at room temperature are shown for the $^{18}$O and $^{16}$O samples (set B) of Na$_x$CoO$_2$ ($x \sim 0.75$). We measured the cleaved surfaces for both of the $^{18}$O (gray solid line) and $^{16}$O (black solid line) samples and the plane surface of a grain only for the $^{18}$O sample (gray dotted line). For the $^{16}$O sample, the two strong phonon modes have been observed at around 465 cm$^{-1}$ and 577 cm$^{-1}$, which can be identified with in-plane $E_{1g}$ and out-of-plane $A_{1g}$ modes of oxygen atoms in the CoO$_6$ octahedra, respectively. For the $^{18}$O sample, the peak positions of these $E_{1g}$ and $A_{1g}$ modes shift to the lower energy side from those of the $^{16}$O sample, respectively.

Fig. 3. (a) Magnetic susceptibilities $\chi$ measured under the ZFC and FC conditions with the magnetic field $H$ = 5 G are shown against $T$ for both the $^{18}$O and $^{16}$O samples of Na$_x$CoO$_2 \cdot y$H$_2$O (set A). (b) The data of $\chi$ around $T_c$ measured under the ZFC condition with $H$ = 5 G are shown against $T$ for both the $^{18}$O and $^{16}$O samples of Na$_x$CoO$_2 \cdot y$H$_2$O (set A). The lines used in the $T_c$-estimation are also shown.

Fig. 4. Magnetic susceptibilities $\chi$ measured with the magnetic field $H$ = 5 G for both the $^{18}$O and $^{16}$O samples of Na$_x$CoO$_2 \cdot y$H$_2$O are shown against $T$ for set B (panel a), set C (panel b) and set D (panel c). The lines used in the $T_c$-estimation are also shown.

Fig. 5. The superconducting transition temperature $T_c$ of Na$_x$CoO$_2 \cdot y$H$_2$O are shown against time $t$ for set A (upper panel) and set C (middle panel). The values of $-4\pi\chi$ observed for the samples are also shown at $T$ = 1.9 K (lower panel). See text for details.

Table I. The superconducting transition temperatures $T_c$ of Na$_x$CoO$_2 \cdot y$H$_2$O are shown for four sets of samples with only $^{16}$O and with ~ 50 % $^{18}$O isotope.

|       | $T_c$ ($^{18}$O) | $T_c$ ($^{16}$O) | $T_c$ ($^{18}$O onset) | $T_c$ ($^{16}$O onset) |
|-------|------------------|------------------|------------------------|------------------------|
| set A | 4.60 K           | 4.58 K           | 4.70 K                 | 4.70 K                 |
| set B | 4.47 K           | 4.48 K           | 4.60 K                 | 4.60 K                 |
| set C | 4.52 K           | 4.49 K           | 4.61 K                 | 4.61 K                 |
| set D | 4.49 K           | 4.52 K           | 4.60 K                 | 4.62 K                 |

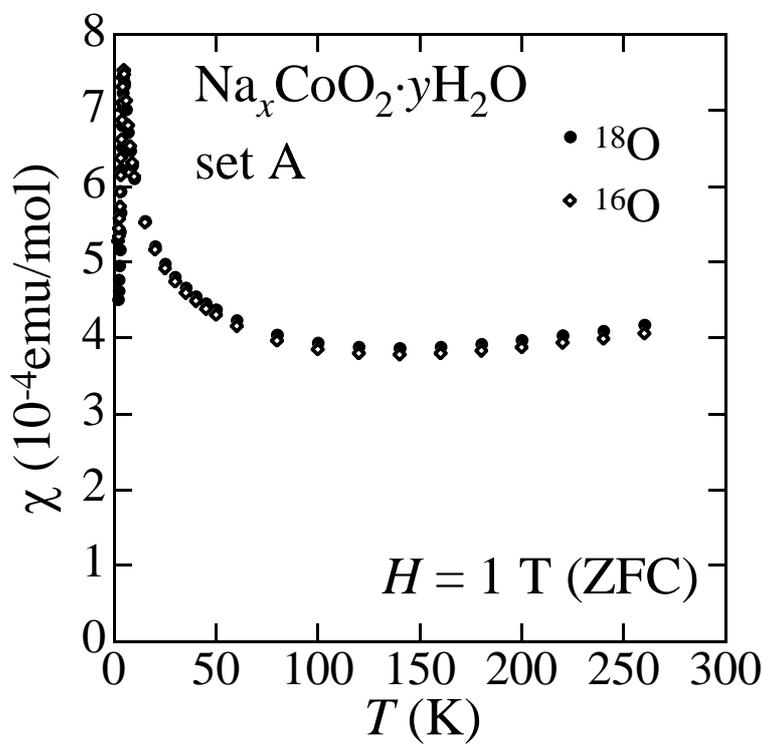

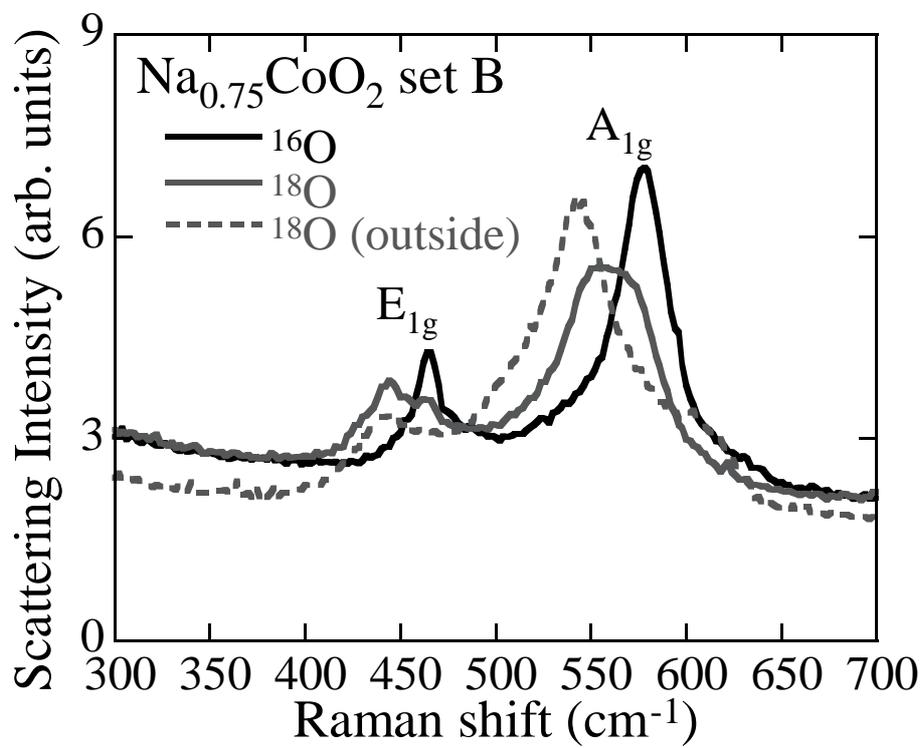

Fig. 2
M. Yokoi *et al.*

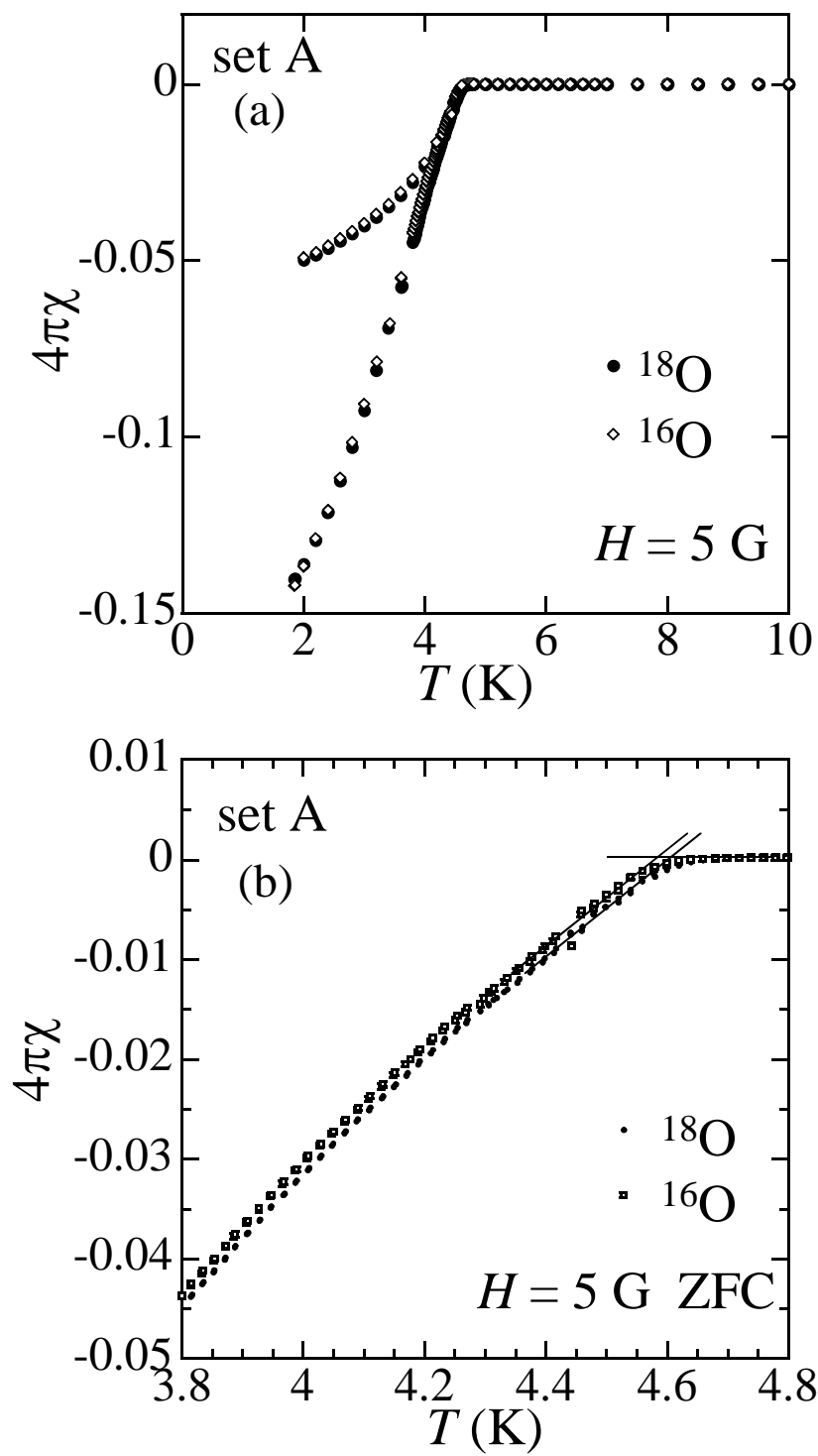

Fig. 3
M. Yokoi et al.

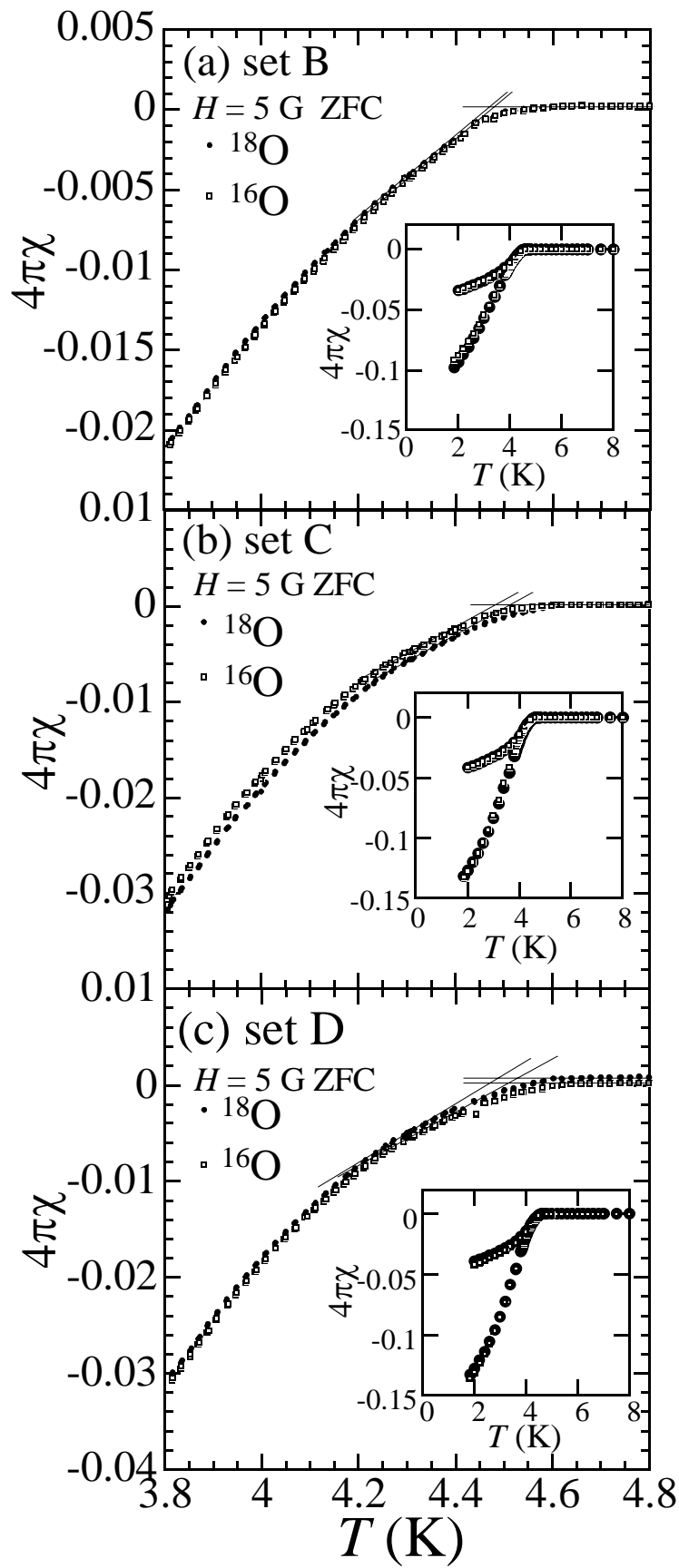

Fig. 4
M. Yokoi *et al.*

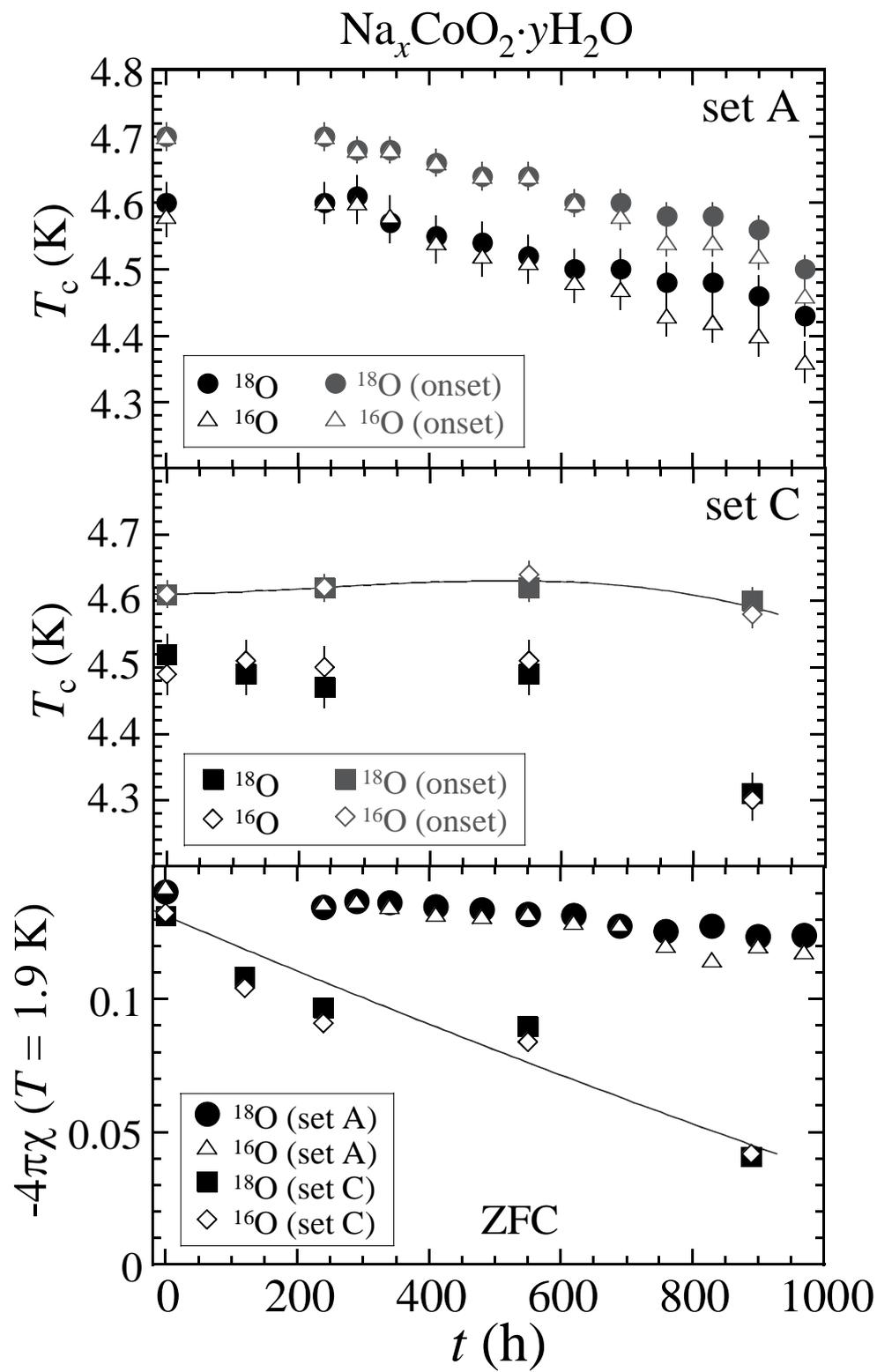

Fig. 5
M. Yokoi et al.